\documentclass[onecolumn,aps,prd,preprintnumbers,superscriptaddress,nofootinbib,amsmath,amssymb,floats,floatfix,showkeys,notitlepage]{revtex4-2}

\usepackage{comment}
\usepackage{lipsum}
\usepackage{graphicx}
\usepackage{subfigure}
\usepackage{palatino}
\usepackage{sans}
\usepackage{hyperref}
\hypersetup{colorlinks=true,linkcolor=blue,urlcolor=blue,citecolor=blue}
\usepackage[toc,page]{appendix}
\usepackage[normalem]{ulem}
\usepackage{adjustbox}
\usepackage{latexsym}
\usepackage{amsmath}
\usepackage{amssymb}
\usepackage{amsfonts}
\usepackage{dcolumn}
\usepackage{bm}
\usepackage{tikz}
\usepackage{bigints}
\usepackage{array,tabularx,multirow,booktabs}
\usepackage[tracking=true]{microtype}
\SetTracking{}{500}
\SetTracking{encoding={*}, shape=sc}{40}
\UseRawInputEncoding %for inputenc error%
\allowdisplaybreaks
\begin{document} \sloppy
\title{Constraining a one-dimensional wave-type gravitational wave parameter through the shadow of M87* via Event Horizon Telescope}

\author{Reggie C. Pantig}
\email{rcpantig@mapua.edu.ph}
\affiliation{Physics Department, Map\'ua University, 658 Muralla St., Intramuros, Manila 1002, Philippines}
\
\begin{abstract}
During the glorious success of the EHT in providing the first image of a black hole, numerous papers have been published about the effect of different astrophysical environments on black hole geometry. Motivated by the work on how gravitational wave affects the shadow of a Schwarzschild black hole [Eur. Phys. J. C 10.1140/epjc/s10052-021-09287-2], we extend it by considering a quantum correction on the black hole through the extended uncertainty principle (EUP). Along with this correction, we probe the gravitational wave's effect on the null geodesics and photonsphere and find constraints to the gravitational wave parameter $\epsilon$ using the black hole shadow of M87* for some given test value for the gravitational wave frequency $\sigma$. Not only were some nodes found in the light trajectory, but the general behavior of paths changes periodically as the time $t$ progresses. These patterns then confirm the chaotic formation of the shadow seen by some remote observer. Finally, the constraint that we find for $\epsilon$ is $~10^{-10}$ orders of magnitude for the effect of the gravitational wave to be seen at a distance of $D = 16.8$ Mpc. As a consequence of such a value for $\epsilon$, another result reveals that while there is are gravitational wave effect on the shadow perceived at $D$, the deviations on the photonsphere are nearly non-existent. Apart from Earth-based detectors for gravitational waves, the study implies the possibility of an alternative detection method, especially if a gravitational wave source is near a lone black hole.
\end{abstract}

%only 5 keywords

\maketitle

%\tableofcontents

\section{Introduction} \label{intr}
As a result of some of the catastrophic, most violent, and energetic processes in the Universe, gravitational waves are produced. The phenomena associated with the tremendous amount of release of energy, which distorts spacetime, can be two massive objects, such as black holes or neutron stars, orbiting each other that eventually merge. Gravitational waves are ripples in spacetime that travel at the speed of light $c$, which were first predicted by Albert Einstein's General Relativity theory (GR) \cite{1916SPAW.......688E,Einstein:1918btx}. It remained a theoretical construct until 2016 when scientists were able to directly detect gravitational waves using the Laser Interferometer Gravitational-Wave Observatory (LIGO) from a binary black hole merger \cite{LIGOScientific:2016aoc}. Such waves are challenging to detect, even with the most sophisticated devices, and are extremely faint. Nonetheless, it opens a new paradigm in astronomy since gravitational waves can be used to probe more subtle information about the Universe than electromagnetic waves. Gravitational waves can provide us with a new way of observing the universe and studying some of the most extreme and exotic objects and phenomena in the cosmos. They offer a unique window into the inner workings of the universe and can help us to understand some of the fundamental laws of nature \cite{LIGOScientific:2018dkp,LIGOScientific:2019fpa,LIGOScientific:2019obb,LIGOScientific:2019ryq,LIGOScientific:2019zcs,DES:2019ccw,Bernal:2020ywq,Lambiase:2020vul,Khodadi:2021ees}.

Another prediction of Einstein's GR is the existence of black holes. The existence of these compact objects, characterized by an extremely strong gravitational field, was verified by the Event Horizon Telescope collaboration by observing the black hole shadows of M87* and Sgr. A* \cite{EventHorizonTelescope:2019dse, EventHorizonTelescope:2022xnr} through electromagnetic means. The dark spot is the silhouette of the event horizon, while the boundary of the black hole shadow, which is invisible in nature, is enveloped by the radiating accretion disk \cite{Dokuchaev:2019jqq}. The black hole shadow with thin accretion disk was first analyzed by Luminet \cite{Luminet:1979nyg}. The shadow contour manifests itself due to the escaped photons under the effect of the small perturbation on the photonsphere radius, which was first studied by Synge \cite{Synge:1966okc}. The results of the EHT were only limited to the confirmation of GR and not to the other alternative theories of gravity. It would take more years, or even more sophisticated equipment, to probe other theories of gravity using the black hole. Ever since many authors have considered exploring the behavior of the classical shadow silhouette for black holes \cite{Dokuchaev:2019jqq,Dokuchaev:2020wqk} described by either a toy model metric or through an alternative theory of gravity \cite{Contreras:2020kgy,Panotopoulos:2021tkk,Panotopoulos:2022bky,Pantig:2022gih,Ovgun:2019jdo,Ovgun:2020gjz,Okyay:2021nnh,Javed:2021arr,Cimdiker:2021cpz,Uniyal:2022vdu,Pantig:2022ely,Rayimbaev:2022hca,Mustafa:2022xod,Pantig:2022qak,Kumaran:2022soh,Atamurotov:2022knb,Atamurotov:2022iwj,Afrin:2021wlj}.

Interestingly, the exploration of the effects of astrophysical environments on the black hole geometry has been recently considered by various authors since one can study these environments indirectly using deviations in the black hole shadow. For instance, the effect of dark matter halo on the black hole shadow was studied \cite{Pantig:2020uhp,Pantig:2022toh,Pantig:2022whj,Pantig:2022sjb,Xu:2020jpv,Xu:2021dkv,Nampalliwar:2021tyz,Jusufi:2020zln,Jusufi:2020cpn,Jusufi:2022jxu,Konoplya:2022hbl,Konoplya:2021ube,Anjum:2023axh,Liu:2022ygf,Khlopov:1985jw}. The possibility of gaining information on the quantum nature of black holes was also considered \cite{Jusufi:2021fek,Devi:2021ctm,Xu:2021xgw,Lobos:2022,Pantig:2021zqe,Chaudhary:2021uuk,Anacleto:2021qoe,Hu:2020usx}. These studies are only a few that we mention from the vast literature, proving the importance of such studies which we cannot underestimate with how fast our space technology evolves.

In this paper, we will consider a black hole that is perturbed by a special type of gravitational wave as an astrophysical environment. Motivated by the paper in \cite{Wang:2019skw}, which explored and studied how the black hole shadow behaves for time-dependent metrics \cite{Wang:2017qhh,Wang:2018eui,Wang:2019tjc,Wang:2022kvg}, we aim to provide an additional analysis by examining the behavior of the null geodesics, as well as find constraints to the gravitational parameter $\epsilon$ by the using the EHT data on M87*. Then, the behavior of the shadow radius will also be studied when it is influenced by different observer locations. To the best of our knowledge, these features are not yet been studied, and the results of this study can add up to the literature concerning gravitational wave effects on the shadow. We will also focus on the non-rotating case for convincing reasons stated in Ref. \cite{Vagnozzi:2022moj}.

The program of the paper is as follows: in the next section, the metric of a perturbed Schwarzschild black hole is introduced. In Sect. \ref{sec3}, the general equations of motion are derived and used to simulate the photonsphere behavior. In Sect. \ref{sec4}, the classical shadow radius is analyzed, and find a constraint to the gravitational parameter $\epsilon$. Such an analysis will only be done on the equatorial plane for brevity. Then, we state concluding remarks and potential research direction in Sect. \ref{conc}. This paper uses natural units such as $G = c = 1$, and the metric signature $(-,+,+,+)$.

\section{Schwarzschild black hole under perturbation} \label{sec2}
The Schwarzschild metric is the simplest black hole solution of the Einstein field equation \cite{Schwarzschild:1916uq}. Recently, a quantum correction through the extended uncertainty principle (EUP), which can be derived from first principles \cite{Costafilho_2016}, was introduced by considering gravitons confined within the event horizon. If $M$ is the geometrized mass of the black hole, then EUP corrected mass is expressed as \cite{Mureika:2018gxl}
\begin{equation} \label{eup}
    \mathcal{M} = M \left(1 + \frac{4\alpha M^2}{L^2_*}\right),
\end{equation}
where $\alpha$ is in unity, and $L_*$ is the large fundamental length scale counterpart of the Planck length $l_p$. In Ref. \cite{Mureika:2018gxl}, the effect of such modification on the mass of the Schwarzschild black hole was studied by the analysis of the deviations through the horizon radius and relativistic orbits of time-like particles. Thereafter, various authors considered exploring the effects of EUP on different black hole phenomena \cite{Lu:2019wfi,Kumaran:2019qqp,Cheng:2019zgc,Hassanabadi:2021kyv,Hamil:2021ilv,Okcu:2022sio,Pantig:2021zqe,Hamil:2022bpd,Chen:2022ngd,Hamil:2020jns,Hamil:2020xud}. Authors in Ref. \cite{Lobos:2022} found constraints to $L_*$ that if uncertainties are ignored, $L_* = 7.950$x$10^{13}$ m for the shadow radius of M87*, $R_\text{sh} = 5.5M$.

A special class of perturbing the Schwarzschild metric to the first order in $\epsilon$ is shown in \cite{Xanthopoulos:1981}, where the line element is expressed as
\begin{equation} \label{metric}
    ds^2 = (g_{\mu\nu} + \epsilon h_{\mu\nu})dx^\mu dx^\nu.
\end{equation}
Here, we now treat Eq. \eqref{eup} to belong in the metric tensors $g_{\mu\nu}$ and $h_{\mu\nu}$. With the perturbation solution considered in \cite{Xanthopoulos:1981}, we rewrite the metric line element as
\begin{equation} \label{metric2}
    ds^2 = -A(t,r,\theta)dt^2 + B(t,r,\theta)dr^2 + C(t,r,\theta)d\theta^2 + D(t,r,\theta)d\phi^2,
\end{equation}
where
\begin{align} \label{metcoef}
    A(t,r,\theta) &= f(r)\left(1 + \epsilon X P_l(\cos\theta) \cos(\sigma t) \right), \nonumber \\
    B(t,r,\theta) &= f(r)^{-1}\left(1 + \epsilon Y P_l(\cos\theta) \cos(\sigma t) \right), \nonumber \\
    C(t,r,\theta) &= r^2\left[1 + \epsilon \left(Z P_l(\cos\theta) + W \frac{d^2}{d\theta^2}P_l(\cos\theta) \right) \cos(\sigma t) \right], \nonumber \\
    D(t,r,\theta) &= r^2 \sin^2(\theta)\left[1 + \epsilon \left(Z P_l(\cos\theta) + W \frac{d}{d\theta}P_l(\cos\theta) \cot(\theta) \right) \cos(\sigma t) \right].
\end{align}
In Eq. \eqref{metcoef}, $P_l$ is the usual Legendre polynomial, and the functions $W, X, Y,$ and $Z$ can be obtained by solving the Einstein field equation $R_{\mu\nu}(g_{\alpha\beta} + \epsilon h_{\alpha\beta}) = 0$ leading to
\begin{align}
    f(r) &= 1 - \frac{2\mathcal{M}}{r}, \quad X = pq, \quad Y = 3\mathcal{M}q, \quad Z = q(r - 3\mathcal{M}), \nonumber \\
    W &= rq, \quad p = \mathcal{M} - \frac{\mathcal{M}^2+\sigma^2 r^4}{r - 2\mathcal{M}}, \quad q = \frac{\sqrt{f(r)}}{r^2}.
\end{align}
The form Eq. \eqref{metric} was first studied in Ref. \cite{Letelier:1996he}, where chaos in time-like orbits are seen through the Melnikov method. More recently, the black hole shadow was studied in Ref. \cite{Wang:2019skw} where chaotic behavior was shown. It is easy to see that without perturbation ($\epsilon = 0$), Eq. \eqref{metric} reduces to the known Schwarzschild metric.

As perturbation theory is so important as a tool for understanding the behavior of black holes and other astronomical objects, it has been used to make predictions about the behavior of black holes under different circumstances, such as analyzing the effects of small perturbations on the background spacetime of a black hole. An example is the presence of gravitational waves or small deviations from the black hole’s equilibrium state. Toward this direction, the stability of the Reisnner-Nordstr\"om black hole was analyzed in Refs. \cite{Zerilli:1974ai,Moncrief:1974gw,Moncrief:1974ng}, leading to the one-dimensional wave-type equation Eq. \eqref{metcoef} when the black hole charge $Q$ vanishes.

\section{Null geodesic} \label{sec3}

For the expressions below, it is understood that the metric functions $A, B, C,$ and $D$ are all functions of the coordinates $t, r,$ and $\theta$.
\begin{align} \label{eos1}
    \mathcal{L},_t &= \frac{1}{2}\left(-A,_{t}{{\dot{t}}}^{2} + B,_{t}{{\dot{r}}}^{2} + C,_{t}{{\dot{\theta}}}^{2} + D,_{t}{{\dot{\phi}}}^{2}\right), \quad \mathcal{L},_{\dot{t}}= -A \dot{t}, \nonumber \\
    \mathcal{L},_r &= \frac{1}{2}\left(-A,_{r}{{\dot{t}}}^{2} + B,_{r}{{\dot{r}}}^{2} + C,_{r}{{\dot{\theta}}}^{2} + D,_{r}{{\dot{\phi}}}^{2}\right), \quad \mathcal{L},_{\dot{r}} = B \dot{r}, \nonumber \\
    \mathcal{L},_\theta &= \frac{1}{2}\left(-A,_{\theta}{{\dot{t}}}^{2} + B,_{\theta}{{\dot{r}}}^{2} + C,_{\theta}{{\dot{\theta}}}^{2} + D,_{\theta}{{\dot{\phi}}}^{2}\right), \quad \mathcal{L},_{\dot{\theta}} = C \dot{\theta}, \nonumber \\
    \mathcal{L},_\phi &= 0, \quad \mathcal{L},_{\dot{\phi}} = D \dot{\phi}.
\end{align}
Indeed, the first line of Eq.\eqref{eos1} indicates the energy per unit mass $E = \mathcal{L},_{\dot{t}}$ is no longer one of the constants of motion. Nevertheless, the fourth line above indicates that there is still spherical symmetry, and $L = \mathcal{L},_{\dot{\phi}}$ is the angular momentum per unit mass of a particle. Note that the coordinates are functions of the affine parameter $\lambda$, which is useful in dealing with null geodesics. Thus,
\begin{align} \label{eos2}
    \frac{d}{d\lambda}\mathcal{L},_{\dot{t}} &= -A\ddot{t}-\left(A,_t\dot{t} + A,_r\dot{r} + A,_\theta\dot{\theta} \right) \dot{t}, \nonumber \\
    \frac{d}{d\lambda}\mathcal{L},_{\dot{r}} &= B\ddot{r}+\left(B,_t\dot{t} + B,_r\dot{r} + B,_\theta\dot{\theta} \right) \dot{r}, \nonumber \\
    \frac{d}{d\lambda}\mathcal{L},_{\dot{\theta}} &= C\ddot{\theta}+\left(C,_t\dot{t} + C,_r\dot{r} + C,_\theta\dot{\theta} \right) \dot{\theta}, \nonumber \\
    \frac{d}{d\lambda}\mathcal{L},_{\dot{\phi}} &= D\ddot{\phi}+\left(D,_t\dot{t} + D,_r\dot{r} + D,_\theta\dot{\theta} \right) \dot{\phi}.
\end{align}
We do not need the fourth expression in Eq. \eqref{eos2} and we use the second expression on the fourth line of Eq. \eqref{eos1}. Provided that the expression for each metric component is quite complicated, we will not write the full expression in this paper.

For null geodesics, it is useful to define $\lambda' = \lambda L$ \cite{Misner:1973prb} (then switch back $\lambda' \rightarrow \lambda$), which then implies that $\mathcal{L},_{\dot{t}} = b^{-1}$ and $\mathcal{L},_{\dot{\phi}} = 1$. Here, the impact parameter is defined as $b = L / E$ where it is understood that $b$ can vary as $t$ varies. Nonetheless, with $ds^2 = 0$, we can obtain an expression for $b$:
\begin{equation} \label{imp}
    b^2 = \left[A \left(C \dot{\theta}^2 + D^{-1}\right)\right]^{-1},
\end{equation}
and thus, we could write
\begin{equation}
    \dot{t} = \left( A b \right)^{-1}.
\end{equation}
With the above expressions, the null geodesics can be obtained and analyzed by using the Euler-Lagrange equation:
\begin{equation} \label{eos3}
    \mathcal{L},_x - \frac{d}{d\lambda}\mathcal{L},_{\dot{x}} = 0.
\end{equation}
\begin{figure*}
    \centering
    \includegraphics[width=0.48\textwidth]{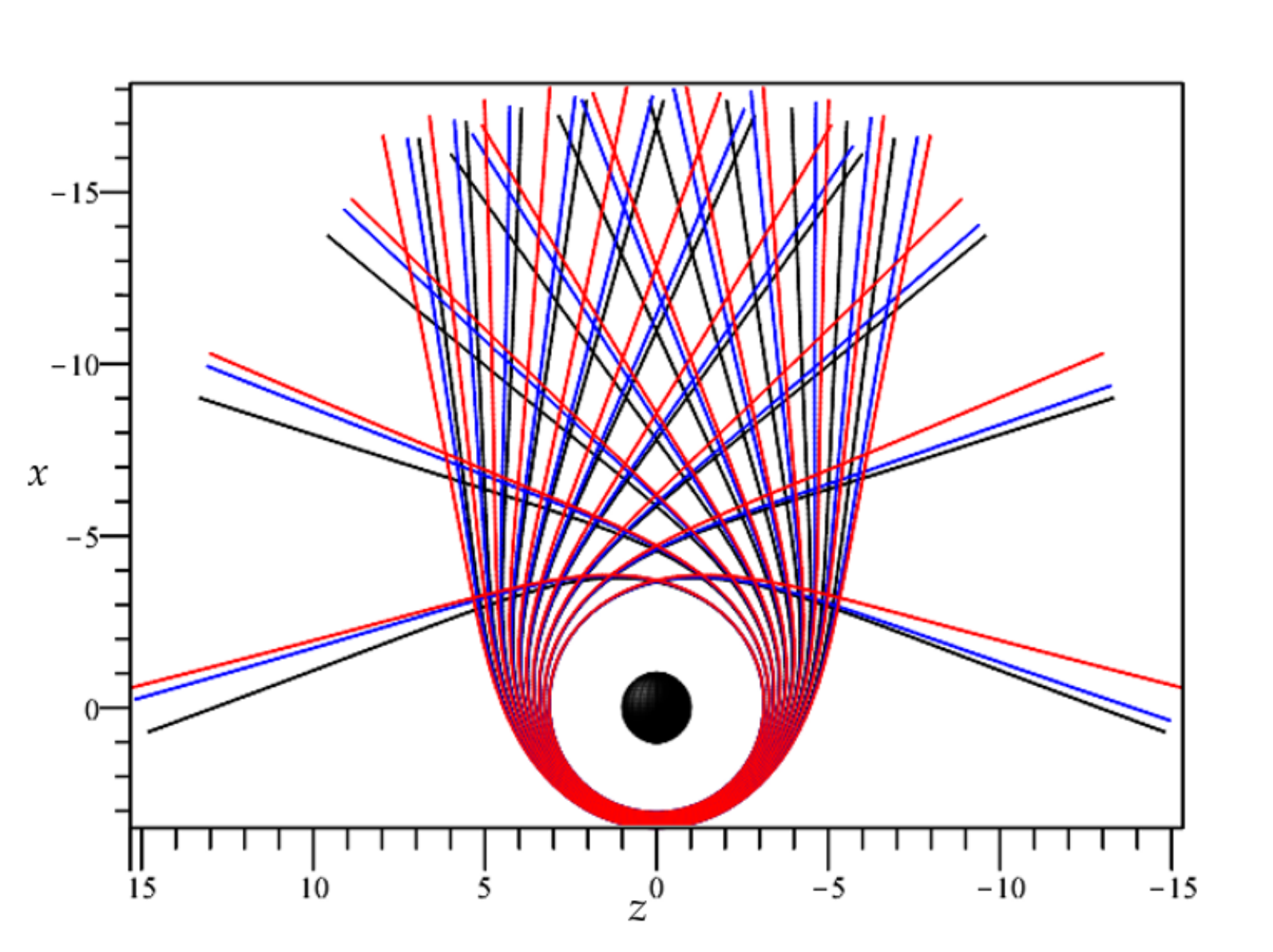}
    \includegraphics[width=0.38\textwidth]{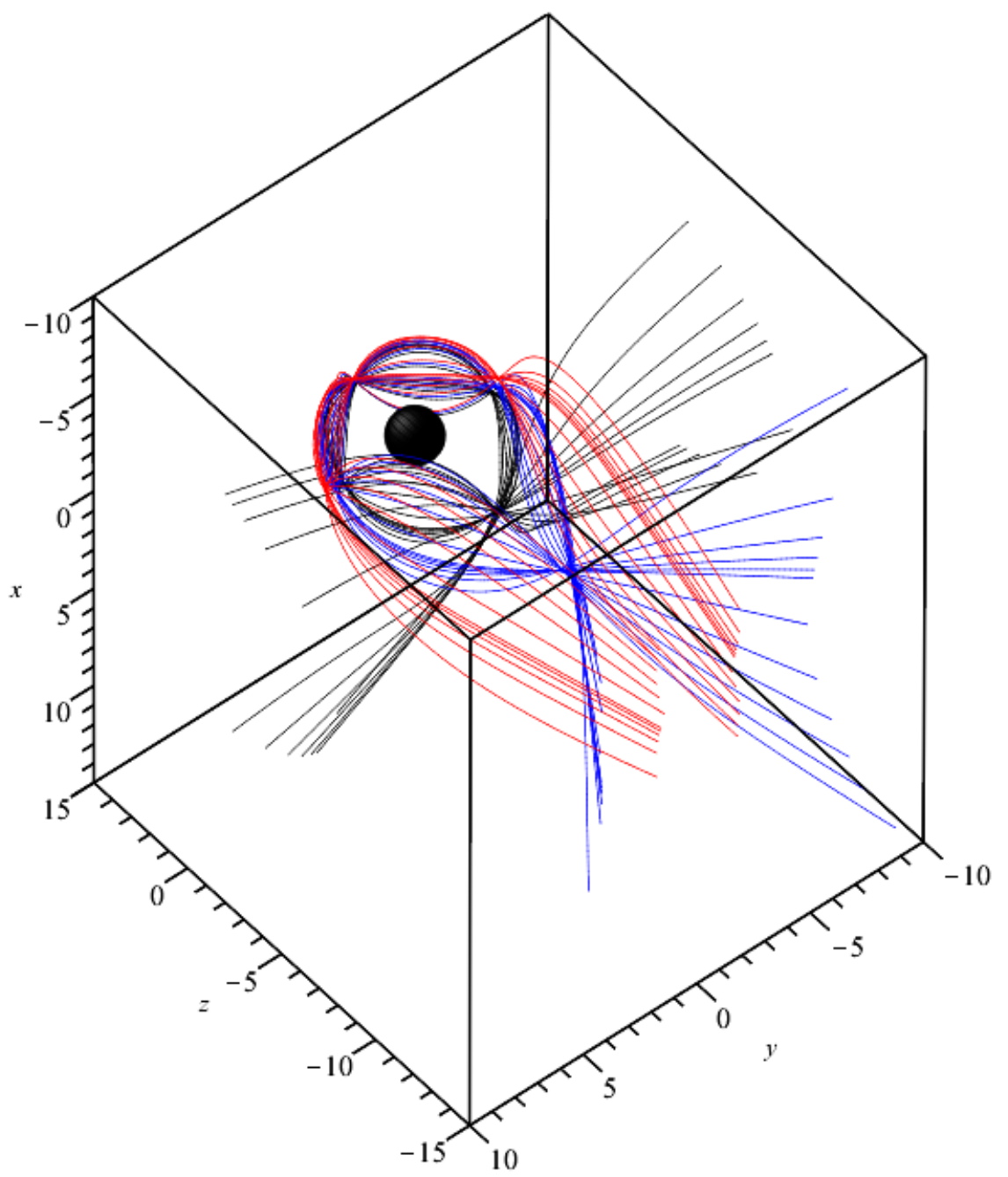}
    \caption{Plots for $l=2$ for $P_l$. The left panel considers $\theta = \pi/2$, while the right panel considers $0 \leq \theta \leq \pi$. The black, blue, and red solid lines correspond to $t=0$, $t=\pi$, and $t=2\pi$, respectively. Furthermore, the chosen values for the parameters $\sigma$ and $\epsilon$ are $0.50$ and $0.05$, respectively.}
    \label{fig:1}
\end{figure*}
While the shadow formation has been studied in Ref. \cite{Wang:2019skw} without analyzing the photonsphere behavior, we present the results of the numerical calculations as shown in Figs. \ref{fig:1}-\ref{fig:3}, where our concern is the case where photons could escape for the possible formation of the shadow. Furthermore, we limit the analysis to excluding photons that escape the event horizon's grip for the sake of simplicity.  In the left panel of these plots, the initial condition corresponds to varying $r$, which is very close to $3M$ for the reason that when $\sigma = 0.50$ and $\epsilon = 0.05$, the photonsphere is $r_\text{ps} \approx 3M$. For this value of $\epsilon$, which is the same value used in Ref. \cite{Wang:2019skw}, the deviation to the Schwarzschild case is only higher by around $10^{-5}$ orders of magnitude. Other initial condition includes $\theta = \pi/2, \dot{\theta} = 0, \dot{t} = 0$, and $\phi = 0$. In the right panel, we fixed $r$ and varied $\theta$. Along the equatorial plane, the effect of the gravitational parameter $\epsilon$ is all present, but the effect of time interval $t$ is rather evident on even orders of $l$ in $P_l$. Finally, when we vary the initial position of the photon in the $\theta$-coordinate, we observe that the photon trajectory becomes more chaotic as $l$ increases. If the backward tracing method is utilized, such as what was done in Ref. \cite{Wang:2019skw}, an observer at $r_\text{o}$ observes chaotic shadow contours.

Notice that the shadow contours in Ref. \cite{Wang:2019skw} were based on the scaling of $r_\text{o}, \sigma,$ and $\epsilon$. It is a procedure implemented in numerical analysis for us to see a quick overview of the phenomenon under study. As an example, the effect of the cosmological constant can be studied by choosing a small value for $\Lambda$, instead of its actual value \cite{Perlick:2018iye, Roy:2020dyy, Maluf:2020kgf, Belhaj:2020kwv, Firouzjaee:2019aij}. Nevertheless, the effect can be seen at low values of the $r-$coordinate instead of actual distances, which is astronomical.  As a final remark, it is interesting that when $\theta$ is varied, we observe some nodal points in the photon trajectories, which also occurs in the photonsphere using a $3$D plot. Furthermore, while the EUP correction increases the size of the photonsphere, the general behavior of the photon trajectories remains the same.
\begin{figure*}
    \centering
    \includegraphics[width=0.48\textwidth]{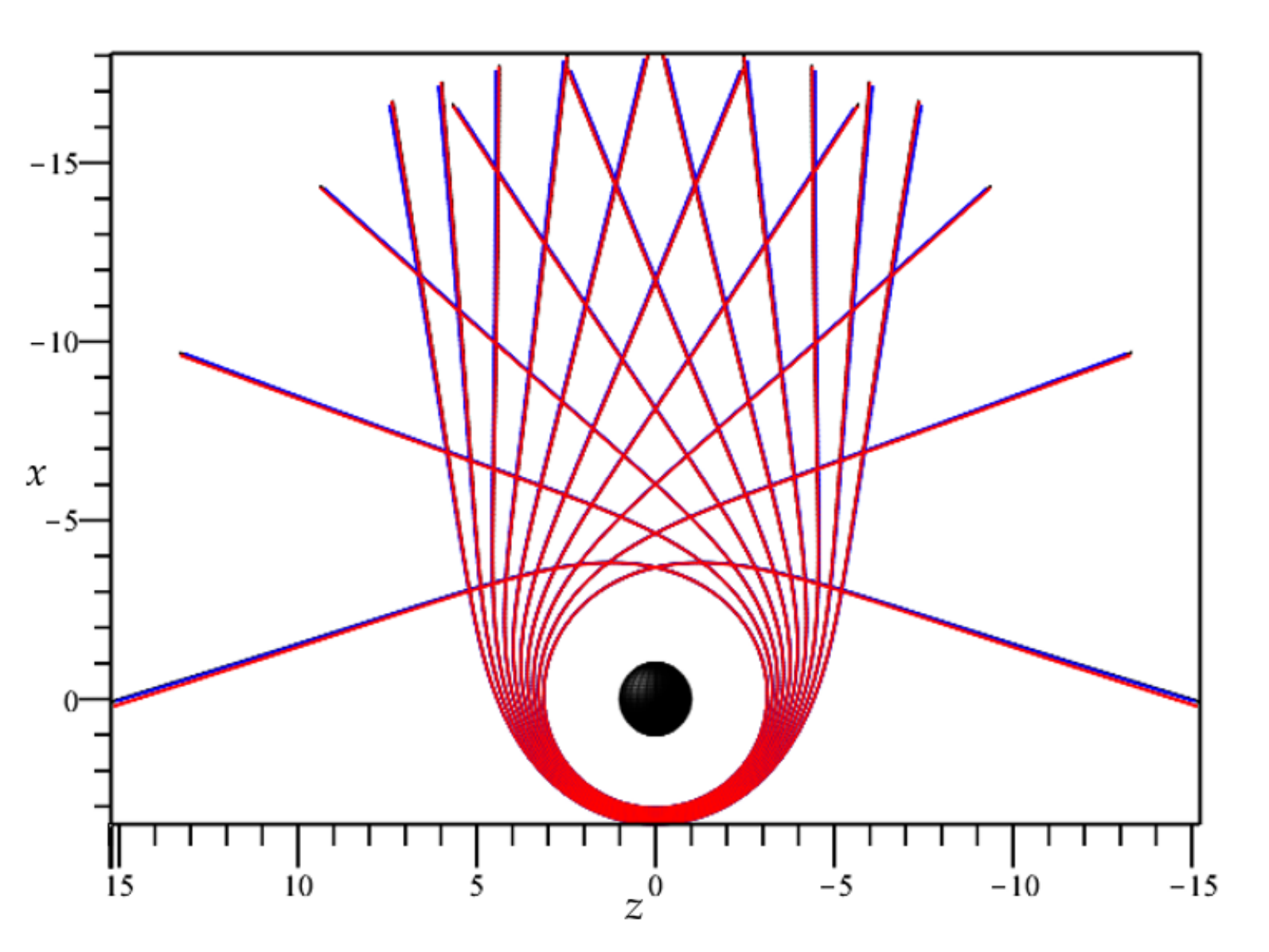}
    \includegraphics[width=0.38\textwidth]{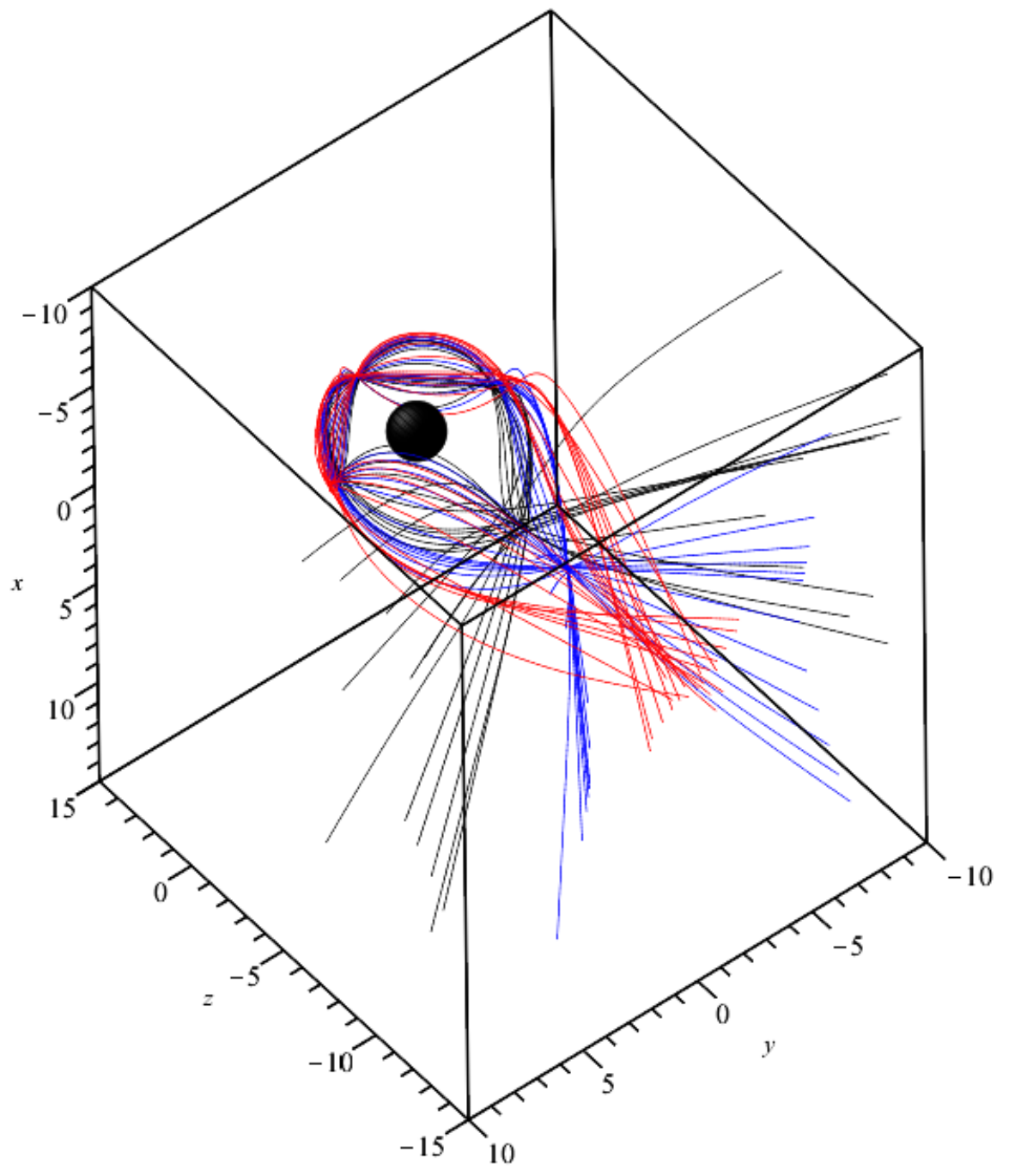}
    \caption{Plots for $l=3$ for $P_l$. The left panel considers $\theta = \pi/2$, while the right panel considers $0 \leq \theta \leq \pi$. The black, blue, and red solid lines correspond to $t=0$, $t=\pi$, and $t=2\pi$, respectively. Furthermore, the chosen values for the parameters $\sigma$ and $\epsilon$ are $0.50$ and $0.05$, respectively.}
    \label{fig:2}
\end{figure*}
\begin{figure*}
    \centering
    \includegraphics[width=0.48\textwidth]{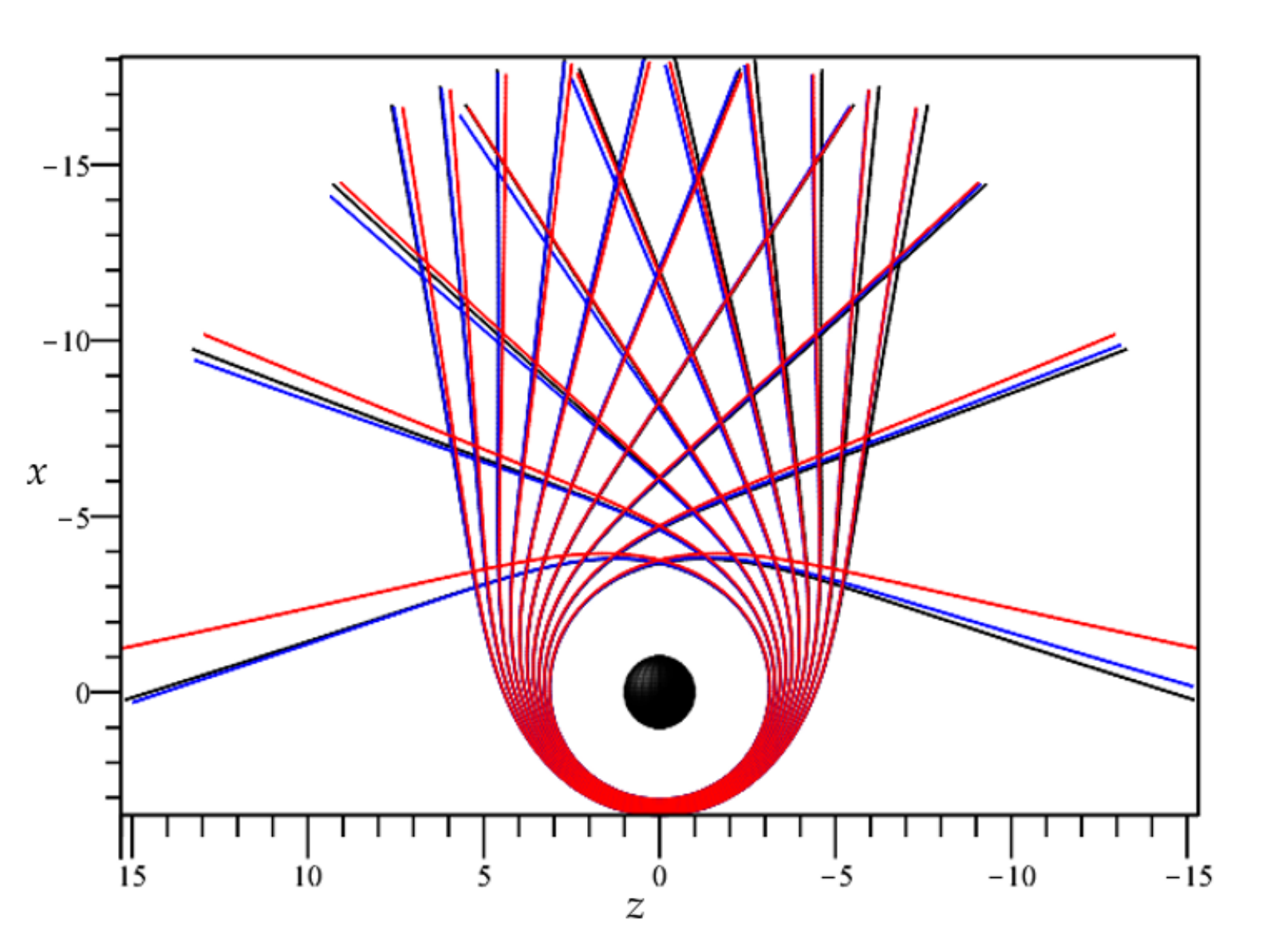}
    \includegraphics[width=0.38\textwidth]{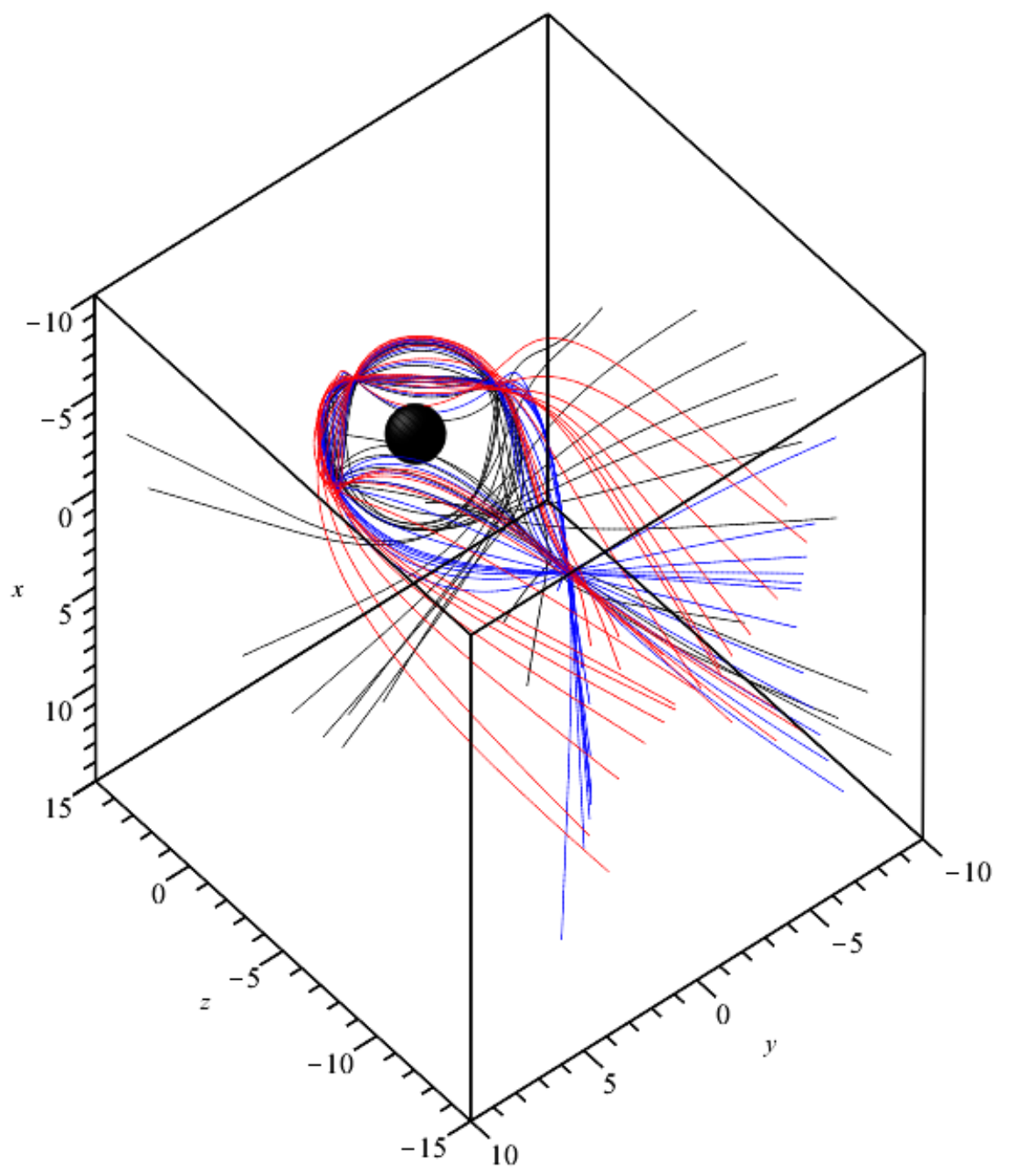}
    \caption{Plots for $l=4$ for $P_l$. The left panel considers $\theta = \pi/2$, while the right panel considers $0 \leq \theta \leq \pi$. The black, blue, and red solid lines correspond to $t=0$, $t=\pi$, and $t=2\pi$, respectively. Furthermore, the chosen values for the parameters $\sigma$ and $\epsilon$ are $0.50$ and $0.05$, respectively.}
    \label{fig:3}
\end{figure*}

Without relying too much on computational power, we can analyze the photonsphere alternatively by specializing in the equatorial plane ($\theta = \pi/2$), which in turn provides us some insight about the shadow behavior at $r_\text{o}$ - the position of some remote observer. These are the spherical photon orbits that under small perturbation could either escape or spiral into the black hole’s gravitational grip. To do so, we will use Hamiltonian formalism. In General Relativity, this is
\begin{equation} \label{ham1}
    H = \frac{1}{2} g^{ik} p_{i} p_{k} = \frac{1}{2} \left( -\frac{p_{t}^{2}}{A} + \frac{p_{r}^{2}}{B} + \frac{p_{\phi }^{2}}{D} \right),
\end{equation}
where we should note that $A, B,$ and $C$ are now only functions of $t$ and $r$.
The equations of motion for null rays are then obtained as follows:
\begin{equation} \label{ham2}
    \dot{x}^{i} = \frac{\partial H}{\partial p_{i}}, \quad \quad \dot{p}_{i} = -\frac{\partial H}{\partial x^{i}},
\end{equation}
where $\dot{x}=dx/d\lambda$ and $\dot{p}$ represents the conjugate momenta. The conjugate momenta give information about some conserved quantity. For instance, executing the operation above gives rise to the following:
\begin{equation} \label{eom1}
    \dot{t} = -\frac{p_{t}}{A}, \quad \quad \dot{r} = \frac{p_{r}}{B}, \quad \quad \dot{\phi } = \frac{p_{\phi }}{D}, 
\end{equation}
\begin{align} \label{eom2}
    \dot{p_t} =& \frac{1}{2}\left(-\frac{p_t^2 A,_t}{A^2} + \frac{p_r^2 B,_t}{B^2} + \frac{p_\phi^2 D,_t}{D^2}\right), \nonumber \\
    \dot{p_r} =& \frac{1}{2}\left(-\frac{p_t^2 A,_r}{A^2} + \frac{p_r^2 B,_r}{B^2} + \frac{p_\phi^2 D,_r}{D^2}\right), \nonumber \\
    \dot{p_\phi} =& 0,
\end{align}
which reveals that there is only a conserved quantity existing along the $\phi$-coordinate, which is the angular momentum per unit mass $p_\phi = L$. It also confirms that there is no more time translational symmetry, since the metric does depend on time coordinate $t$, and the energy $E = -p_t$ is no longer conserved.

Looking at Eq. \eqref{eom1}, it does not restrict us to write
\begin{equation} \label{imp1}
    \frac{L}{E} = \frac{D}{A} \frac{d\phi}{dt},
\end{equation}
which is just the impact parameter $b$, equal to Eq. \eqref{imp}. It is clear that $b$ now depends on $r$ and $t$. Then, if one specifies a certain value for $t$, there would be a certain value for $b$ along with the given $r$. If $r$ varies as $t$ remains constant in the impact parameter, then we have
\begin{equation} \label{imp2}
    b,_r  = \frac{AD,_r - DA,_r}{A^2}
\end{equation}

\begin{figure*}
    \centering
    \includegraphics[width=0.48\textwidth]{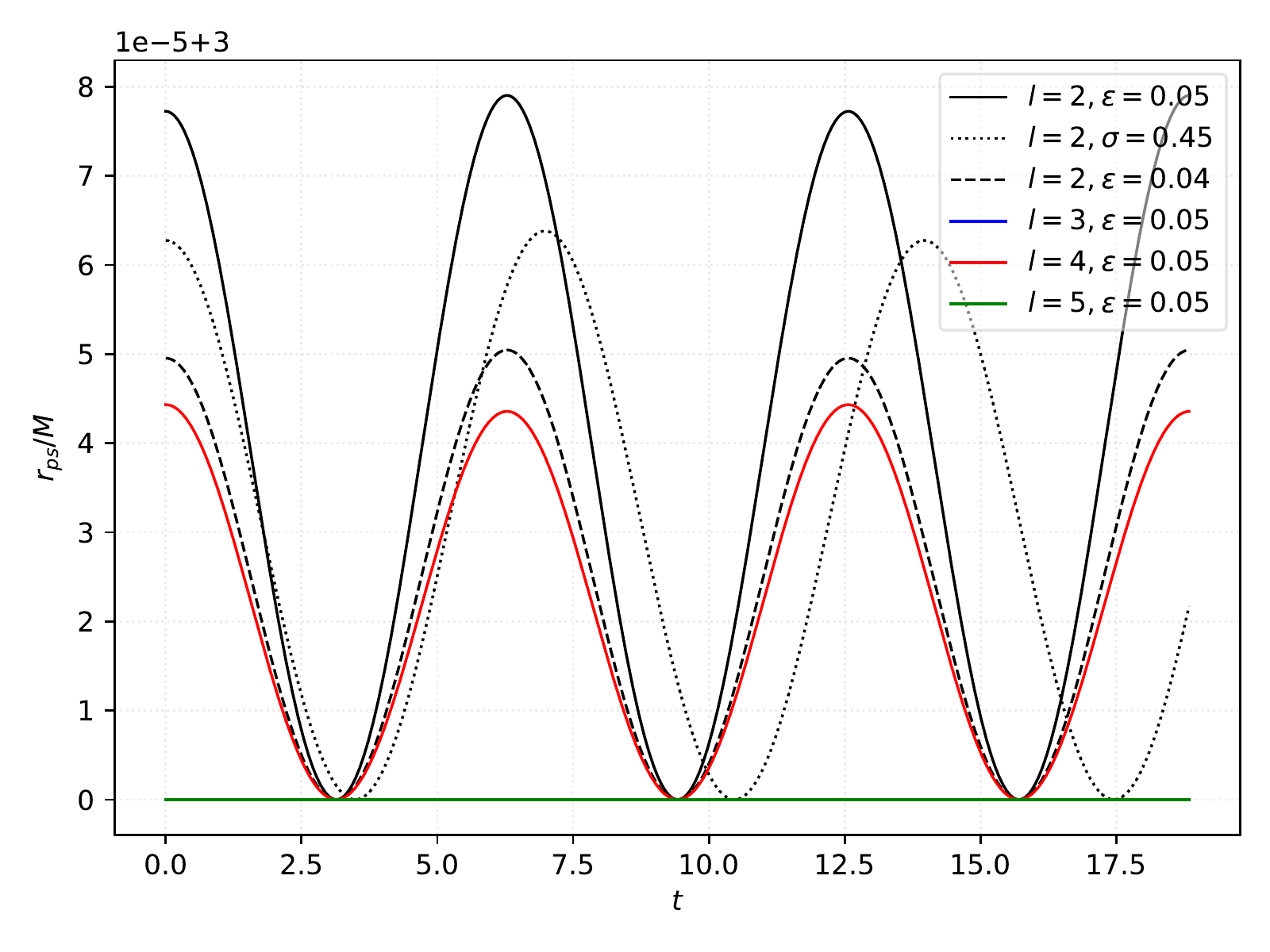}
    \includegraphics[width=0.48\textwidth]{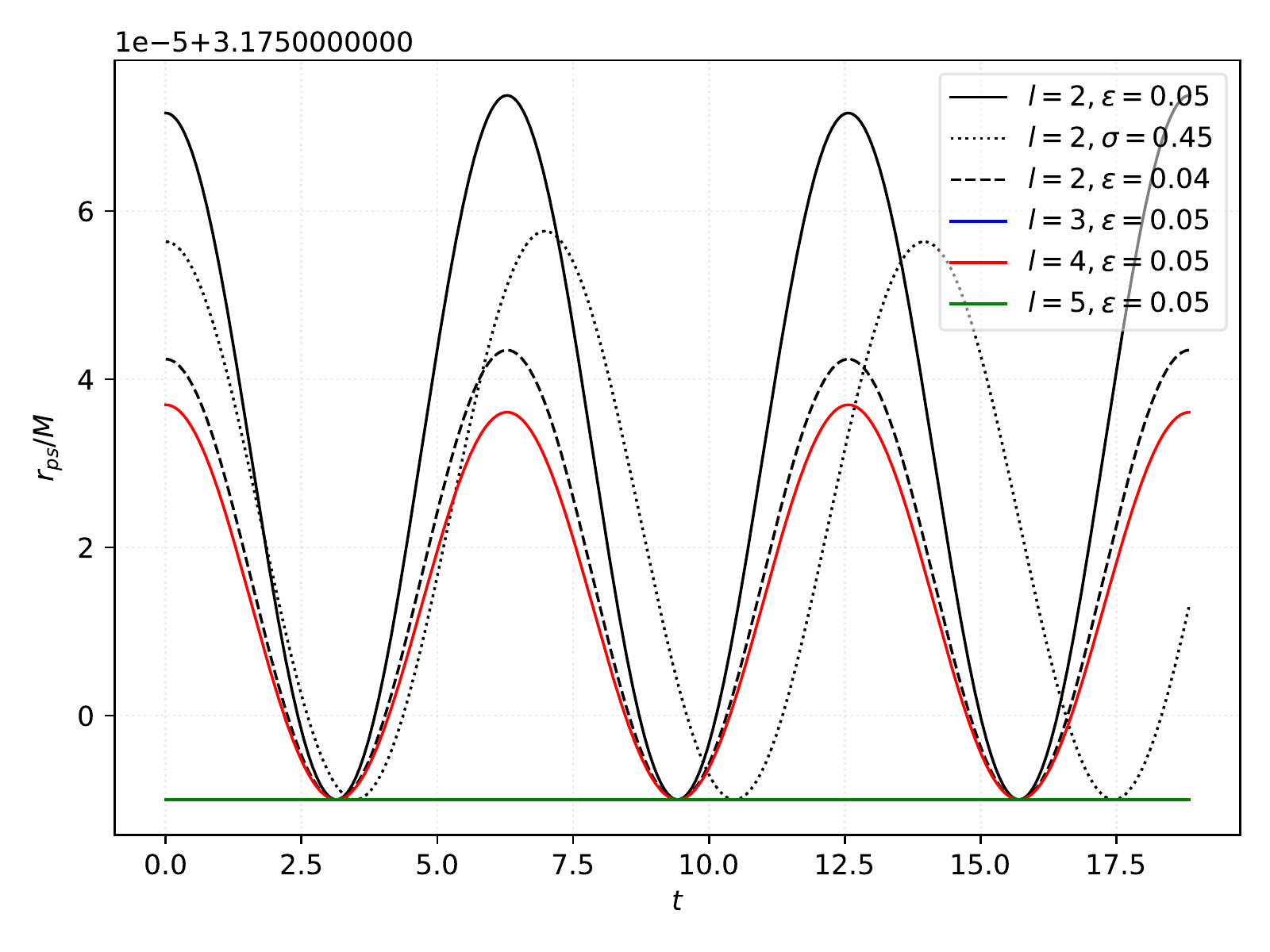}
    \caption{Behavior of the photonsphere radius at different time $t$ for a given value of gravitational wave frequency $\sigma = 0.50$ along a fixed polar angle $\theta = \pi/2$ for an observer. The left panel is without the EUP correction. In the right panel, we have chosen M87* where $L_* = 7.95$x$10^{13}$ m.}
    \label{fig_rps}
\end{figure*}

Next, from Eq. \eqref{eom1}, we obtain
\begin{equation} \label{eqorb}
    \frac{\dot{r}}{\dot{\phi }} = \frac{dr}{d\phi } = \frac{D} {B}\frac{p_r}{p_{\phi}},
\end{equation}
which can be combined to $H = 0$ in Eq. \eqref{ham1} resulting to
\begin{equation} \label{eorb}
    \frac{dr}{d\phi } = \left[ {\frac {D  }{B  } \left( {\frac {D }{A  {b}^{2}}}-1 \right) }\right]^{1/2}.
\end{equation}
The radius of the photonsphere can be obtained by setting $dr/d\phi = 0$, and $d^2r/d \phi ^2 =0$. The second condition gives the expression
\begin{equation}
    A b D_,r  -2 A D b,_r - D b A,_r  = 0
\end{equation}
by which using Eqs. \eqref{imp} and \eqref{imp2}, we obtain a way to extract the photonsphere through
\begin{equation}
    AD,_r - DA,_r = 0,
\end{equation}
which is reminiscent to the standard formula without time dependence $(D'(r) A(r) - D(r)A'(r) = 0)$, where the prime denotes derivative with respect to $r$. In Fig. \ref{fig_rps}, we numerically plot the different values of the photonsphere radius $r_\text{ps}$. We observe an oscillatory behavior for even order $l$ in $P_l$ with the lower limit of $r_\text{ps} = 3M$ - the Schwarzschild case. However, for the odd order of $l$ in $P_l$, the behavior of the photonsphere coincides with that of the Schwarzschild case. For comparison, the gravitational effect decreases as we decrease the parameter $\epsilon$'s value. We can say that such a parameter dictates the magnitude of the deviation from the Schwarzschild case. We considered both the gravitational wave and EUP effects in the right panel. We see that the lower limit of the photon radius increases ($r_\text{ps} = 3.175M$) due to the EUP correction as the gravitational wave effect occurs above it. We also remark that the value of the observed peak cycles every two consecutive wavelengths. In addition, when we decrease $\sigma$, the amplitude, and wavelength of the gravitational wave decrease and increase, respectively.

\section{Equatorial constraints in $\epsilon$ through the black hole shadow} \label{sec4}
In this section, we aim to find constraints $\epsilon$ using the EHT data of the black hole shadow in a simplified case - at the equatorial plane only through different slices of $t-$coordinate. Thus, we expect to find different values of $\epsilon$ due to the time dependence of the shadow. If there exists a closed formula for the shadow at varying $\theta$, we may deduce that will also be the case, and the only important information is to get the behavior of the bounds of $\epsilon$ and how small it is. At the same time, there is available data for Sgr. A* and M87* about the SMBH's angular radius (see Table \ref{tab1}), we will only use M87* in this study. One can calculate the shadow diameter using $d_\text{sh} = D \alpha_\text{sh} / M$, which results to $d^\text{M87*}_\text{sh} = (11 \pm 1.5)m$. We present two caveats before proceeding; First, a gravitational wave is assumed to pass through M87*. Second, we can assume a non-spinning case due to the convincing reasons stated in Ref. \cite{Vagnozzi:2022moj} (see page $5$).

Consider a stationary observer situated at the coordinates $(t, r, \theta, \phi) = (t_\text{o},r_\text{o},\theta_o = \pi/2, \phi_o = 0)$. At this specific condition, the observer can define a function that gives information about the angular radius of the shadow \cite{Perlick:2015vta}:
\begin{equation}
    \tan(\alpha_{\text{sh}}) = \lim_{\Delta x \to 0}\frac{\Delta y}{\Delta x} = \left(\frac{D}{B}\right)^{1/2} \frac{d\phi}{dr} \bigg|_{r=r_\text{o}}.
\end{equation}
Using Eq. \eqref{eorb}, we can rewrite the above equation as
\begin{equation}
    \sin^{2}(\alpha_\text{sh}) = \frac{b_\text{crit}^{2}(r_\text{ps},t_\text{o})}{h(r_\text{o},t_\text{o})^{2}}.
\end{equation}
In the above expression, the critical impact parameter can be obtained by the condition that $\frac{d^2r}{d\phi^2} = 0$, which gives
\begin{equation}
    b^2_\text{crit} = {\frac {D \left( r,t_\text{o} \right)  \left( 2\, \left( {\frac {\partial }{\partial r}}D \left( r,t_\text{o} \right)  \right) A \left( r,t_\text{o} \right) B\left( r,t_\text{o} \right) -D \left( r,t_\text{o} \right)  \left( {\frac {\partial }{\partial r}}B \left( r,t_\text{o} \right)  \right) A \left( r,t_\text{o} \right) -B\left( r,t_\text{o} \right) D \left( r,t_\text{o} \right) {\frac {\partial }{\partial r}}A \left( r,t_\text{o} \right)  \right) }{ \left( A \left( r,t_\text{o} \right) \right) ^{2} \left(  \left( {\frac {\partial }{\partial r}}D \left( r,t_\text{o} \right)  \right) B \left( r,t_\text{o} \right) -D \left( r,t_\text{o} \right) {\frac {\partial }{\partial r}}B \left( r,t_\text{o} \right)  \right) }} \bigg|_{r=r_\text{ps}},
\end{equation}
and $h(r_\text{o},t_\text{o})$ is a defined by
\begin{equation}
    h(r_\text{o},t_\text{o}) = \sqrt{\frac{D(r_\text{o},t_\text{o})}{A(r_\text{o},t_\text{o})}}.
\end{equation}
Finally, the shadow radius can be calculated through
\begin{equation} \label{eshad}
    R_\text{sh} = r_\text{o} b_\text{crit}(r_\text{ps},t_\text{o}) \sqrt{\frac{A(r_\text{o},t_\text{o})}{D(r_\text{o},t_\text{o})}}.
\end{equation}

\begin{figure*}
    \centering
    \includegraphics[width=0.48\textwidth]{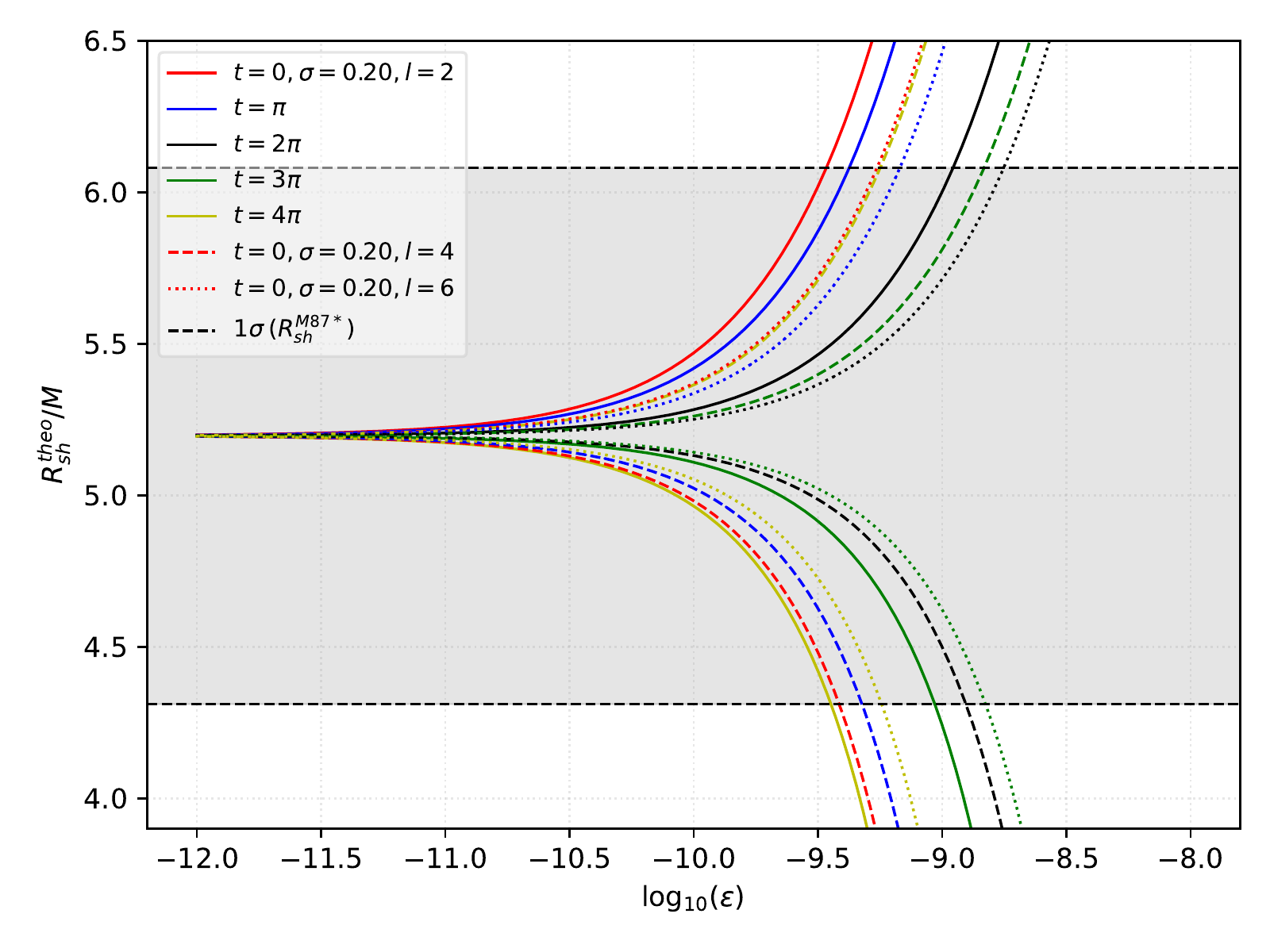}
    \includegraphics[width=0.48\textwidth]{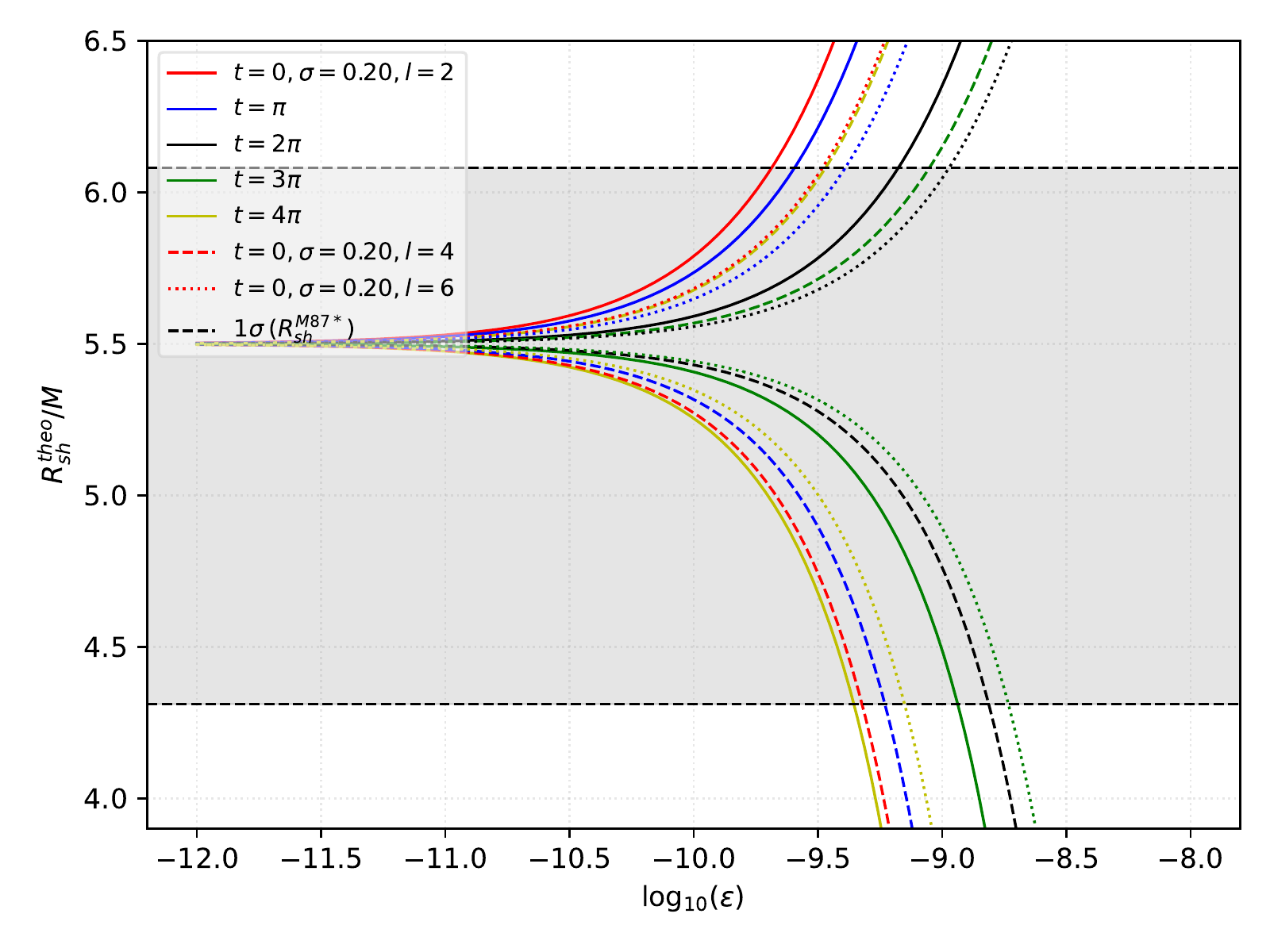}
    \caption{Left panel: without EUP correction. Right panel: with EUP correction. The plots constrain the value of $\epsilon$ using the EHT data for M87*'s shadow.}
    \label{fig2}
\end{figure*}

The result is shown in Fig. \ref{fig2}. Here, we used the allowed  $1\sigma-$ bands for the Schwarzschild deviation \cite{EventHorizonTelescope:2019dse,EventHorizonTelescope:2022xnr,EventHorizonTelescope:2021dqv,Vagnozzi:2022moj}, which read $ 4.31M \leq R_\text{sh} \leq 6.08M$ for M87*. In the left panel, there is no EUP correction and the SMBH is described by the perturbed Schwarzschild black hole. In constraining $\epsilon$, we considered a test frequency for the gravitational wave, which is $\sigma = 0.20$. Each curve represents a different time-coordinate $t$, and the curve type corresponds to a different order in $l$. We can see that as $\epsilon \rightarrow 0$, the oscillatory behavior of the shadow can be nearly neglected. However, near the bounds, considerable deviations can be observed as $t$ and $l$ are changed. As for the right panel, a close inspection shows that the bounds $\epsilon$ become smaller compared to the case where EUP correction is not considered.
\begin{table}
    \centering
    \begin{tabular}{ p{2cm} p{3.5cm} p{4.5cm} p{2cm}}
    \hline
    \hline
    Black hole & Mass $m$ ($M_\odot$) & Angular diameter: $2\alpha_\text{sh}$ ($\mu$as) & Distance $D$ (kpc) \\
    \hline
    Sgr. A*   & $4.3 \pm 0.013$x$10^6$ (VLTI)    & $48.7 \pm 7$ (EHT) &   $8.277 \pm 0.033$ \\
    M87* &   $6.5 \pm 0.90$x$10^9$  & $42 \pm 3$   & $16800$ \\
    \hline
    \end{tabular}
    \caption{Black hole observational constraints.}
    \label{tab1}
\end{table}

These plots tell us how the shadow radius behaves due to the values of $\epsilon$ for a given $\sigma$. We observe that as time differs by $n\pi$ where $n$ is an integer, the black hole shadow increases or decreases in radius if $n$ is small. We also verified that as $n \rightarrow \infty$, the curve follows the Schwarzschild trend, and oscillation ceases to exist. We also emphasize that the plot does not reveal the shape of the shadow silhouette (for that, see Ref. \cite{Wang:2019skw}). Nevertheless, even if the shadow contour is distorted to any shape (like the case for the Kerr shadow which is D-shaped), the calculation for $R_\text{sh}$ can still be accomplished. Some final remarks. We saw that the effect of the very small value of $\epsilon$ is to observe the cyclical increase and decrease of the shadow radius. However, its corresponding effect on the photonsphere radius is vanishingly small. Such a finding indicates that even a very small perturbation can have a long-term effect on photons as it travels through space and reach the remote observer.
\begin{figure*}[!ht]
    \centering
    \includegraphics[width=0.48\textwidth]{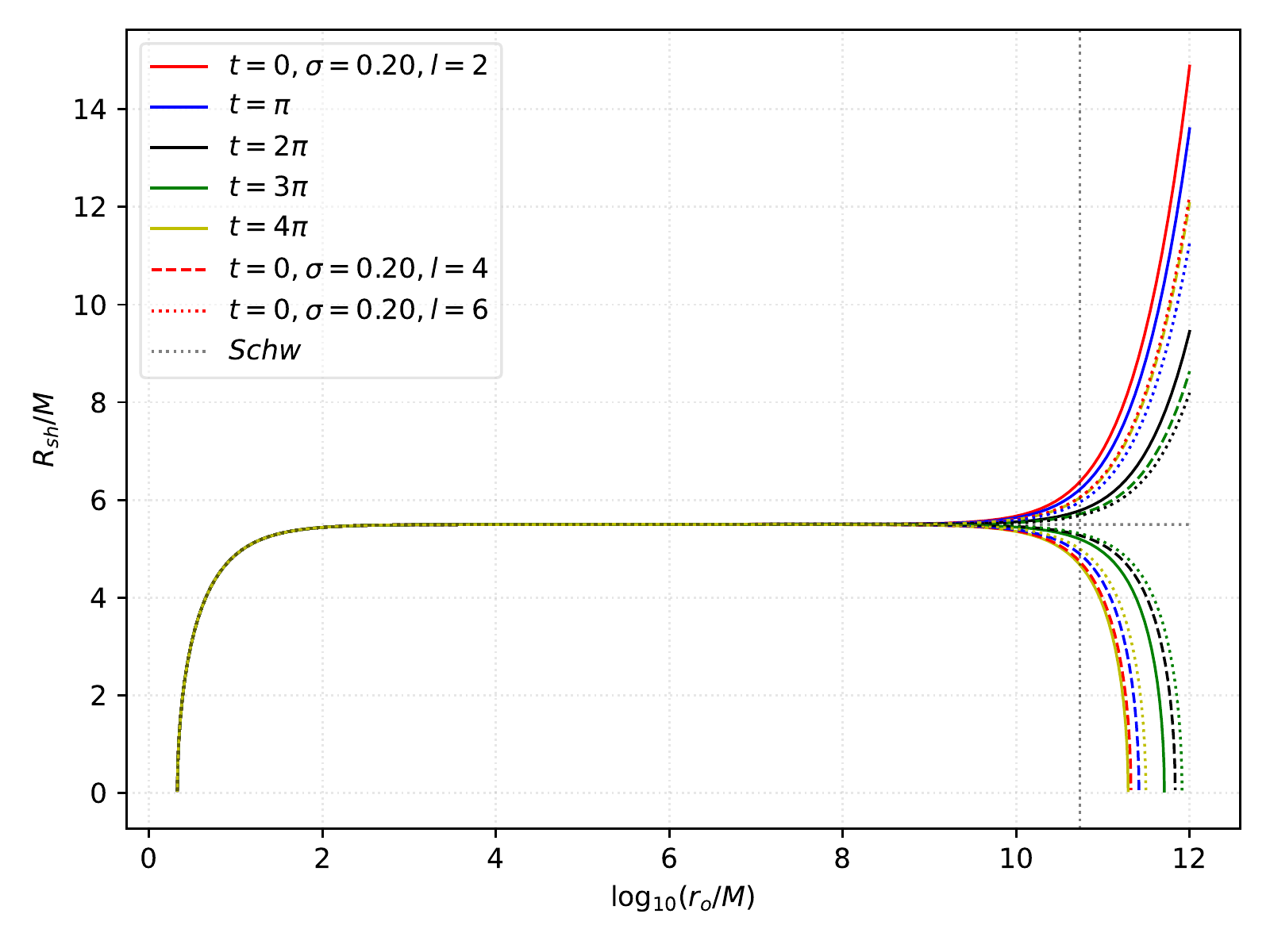}
    \includegraphics[width=0.48\textwidth]{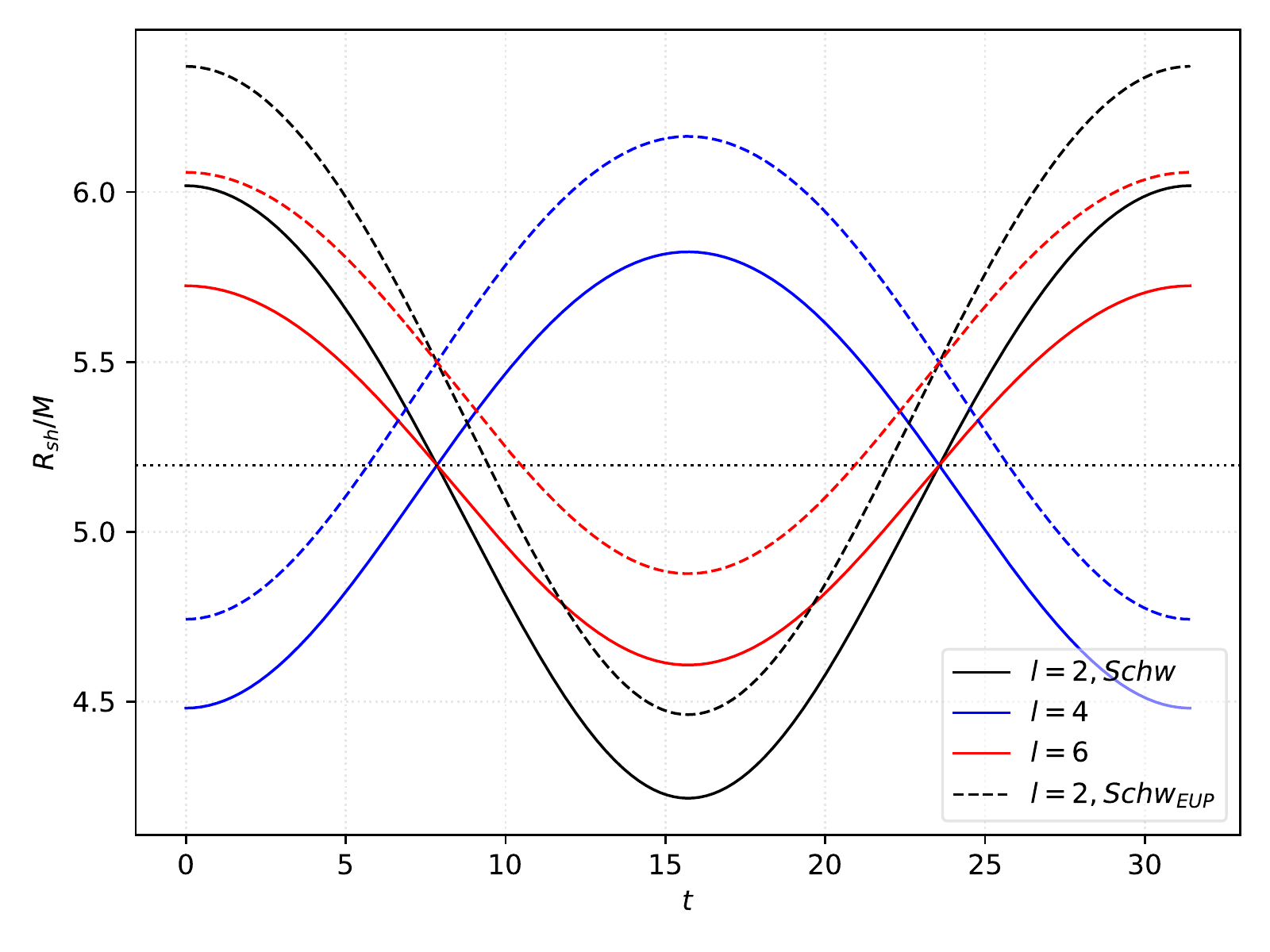}
    \caption{Left: Shadow radius behavior as the $r_\text{o}$ varies. Right: Shadow radius behavior with EUP correction as the time-coordinate $t$ varies. The location of the observer is fixed at $r_\text{o} = 16.8$ Mpc.  In these plots, $\sigma = 0.20$, $\epsilon = 10^{-9.5}$.}
    \label{fig3}
\end{figure*}

Next, let us consider choosing an estimated value for the gravitational wave parameter $\epsilon = 10^{-9.5}$ based on Fig. \ref{fig2}. Then, to consider the effect of varying $r_\text{o}$ on the black hole shadow under the influence of the gravitational wave, we plot Eq. \eqref{eshad} again, placing $r_\text{o}/M$ along the horizontal axis. See Fig. \ref{fig3}, where the vertical dotted line corresponds to $r_\text{o} = 16.8$. The plot on the right visualizes that the effect of the gravitational wave on the shadow radius is negligible to none at locations $\log_{10}(r_\text{o}/M) \leq 9$. In the figure to the right, we can see the oscillatory behavior of the shadow radius as time goes by for different even orders of $l$ in $P_l$. The dotted horizontal line corresponds to the Schwarzschild case where $R_\text{sh}=3\sqrt{3}M$. Meanwhile, we can see that the EUP effect merely increases the shadow radius relative to the one without EUP correction.

\section{Conclusion} \label{conc}
In this work, we first explored the null geodesics of a black hole with quantum correction as perturbed by a gravitational wave which came from a special class of a $1$D wave-type equation. To do so, we derived the geodesic equation used for numerical considerations, which allows for examining the behavior of the photon trajectory through different orders of $l$ in $P_l$. We have seen that at the equatorial plane, the effect of the gravitational wave parameter $\epsilon$ is strong for even orders of $l$. Nevertheless, the full behavior of the photon trajectories can be seen as one considers varying the initial points in $\theta$. It reveals that as $l$ increases, the photon trajectories become more chaotic along with the progress of time $t$. An immediate implication is the chaotic shadow contours, which were analyzed in Ref. \cite{Wang:2019skw}. With the EUP correction, both the photonsphere and shadow radius are affected and that is to increase its value relative to the Schwarzschild case.

We also analyzed the behavior of photonsphere radius along the equatorial plane for the sake of brevity, and we found that deviations only occur at even orders of $l$ in $P_l$. Such a deviation, which decreases as $l$ increases, is very small relative to the Schwarzschild case given that $\epsilon = 0.05$. There is some dual effect when it comes to the gravitational wave frequency and that is to both decrease the peak values of the deviation and its time of occurrence. Finally, we constrained the value of $\epsilon$ through the shadow observation on M87* by EHT, limiting the situation only along the equatorial plane at different slices of $t-$coordinate. This is shown in Fig. \ref{fig2}, where the constraints change through time relative to point of observation $t_\text{o}=0$. Given the parameters, we can predict that as $l$ increases, the upper range for $\epsilon$ also increases. We remark that the range for $\epsilon$ changes, for a gravitational wave of different frequency $\sigma$. Although we assumed theoretically the notion that a gravitational wave passes through M87*, the important aspect of this study is to reveal the possible range of the orders of magnitude of $\epsilon$ given $\sigma$. We know that this should also happen if different $\theta_\text{o}$ is considered. At the moment, the program used in Ref. \cite{Wang:2019skw} to generate the shadow contour in detail cannot be used for constraining since the results came from scaling $r_\text{o}$ and $\epsilon$.

While it is true that $h_{\mu\nu}$ is divergent as $r \rightarrow \infty$ \cite{Wang:2019skw}, it was shown in this paper with our finite distance from M87*, although considerably vast, the possibility of observing the effect of a gravitational wave through the black hole shadow is still in at hand provided a proper choice for the gravitational wave parameter $\epsilon$, which must be treated as very small. It can be verified that if $\epsilon \rightarrow 0$, similar to the estimate used in Fig. \ref{fig3}, the gravitational wave's effect on the photonsphere radius is vanishingly small, in contrast to Figs. \ref{fig:1}-\ref{fig:3} where $\epsilon$ is called so that the gravitational effects can be observed at low values of $r_\text{o}$. With the constraint on $\epsilon$ it is found that the gravitational wave effect on the shadow does not show up as the observer becomes close to the perturbed black hole. Indeed, it confirms that the perturb metric associated with this special class of gravitational wave is not asymptotically flat, as mentioned in Ref. \cite{Wang:2019skw} and its effect is revealed at a vast distance from the black hole. The EUP effect, however, is found to further enhance the enlargement of the shadow as an observer gets so far away.

Finally, we indicate some research directions. It would be interesting to explore the effect of charge $Q$ in the analysis since originally, Eq. \eqref{metcoef} came from a Reissner-Nordst\"om solution. Perhaps a more challenging work is to consider the effect of a gravitational wave on the time-like, null geodesic, and the shadow of a Kerr black hole since it offers a more realistic scenario. Another prospect is to consider other types of gravitational waves, and one may consider analyzing their effect on the shadow and deflection angle in the strong and weak field regimes.

\begin{acknowledgements}
R. P. would like to acknowledge networking support by the COST Action CA18108 - Quantum gravity phenomenology in the multi-messenger approach (QG-MM).
\end{acknowledgements}

\bibliography{ref}
\bibliographystyle{apsrev}
\end{document}